\documentstyle[aps,prb]{revtex}
\begin{document}
\twocolumn[{
\draft
\widetext

\title{\bf How Phase-Breaking Affects Quantum Transport
Through Chaotic Cavities}

\author{Harold U. Baranger$^{1}$ and Pier A. Mello$^{2}$}

\address{$^{1}$ AT\&T Bell Laboratories 1D-230, 600 Mountain Avenue,
Murray Hill, New Jersey 07974-0636}
\address{$^{2}$ Instituto de F\'{\i}sica, Universidad Nacional Aut\'{o}noma
de M\'{e}xico, 01000 M\'{e}xico D.F., M\'{e}xico}

\date{Submitted to Phys. Rev. B, November 14, 1994}
\maketitle

\mediumtext
\begin{abstract}
We investigate the effects of phase-breaking events on electronic
transport through ballistic chaotic cavities. We simulate phase-breaking
by a fictitious lead connecting the cavity to a phase-randomizing
reservoir and introduce a statistical description for the total scattering
matrix, including the additional lead.
For strong phase-breaking, the average and variance of the conductance are
calculated analytically.  Combining these results with those in the
absence of phase-breaking, we propose an
interpolation formula, show that it is an excellent description of
random-matrix numerical calculations, and obtain good agreement with
several recent experiments.
\end{abstract}

\pacs{PACS numbers: 72.20.My, 05.45.+b, 72.15.Gd}
}]

\narrowtext

Recently there has been great interest in the
effects of quantum-mechanical interference on electronic transport through
ballistic quantum dots.\cite{ReviewMes} In these microstructures both
the phase-coherence length and the elastic mean free path
exceed the system dimensions.  Thus the leads into the dot can be
thought of as electron waveguides and the dot itself as a resonant cavity.

Experimentally\cite{Mar92,Mar93,MJBer94,Kel94,Weiss93,Schuster94,Chang94,chan}
one observes random (but reproducible) fluctuations in the conductance as the
magnetic field,\cite{Mar92,MJBer94,chan} the Fermi energy,\cite{Kel94} or
the shape of the cavity\cite{chan} is changed.
The sensitivity to small changes in these parameters shows that these
fluctuations are caused by quantum interference. In addition to conductance
fluctuations, several interference effects which survive averaging over
many microstructures are observed;\cite{Weiss93,Schuster94,Chang94}
in particular, an increase in the average resistance at zero magnetic field
called the weak-localization correction (WLC).\cite{Chang94} The two main
features of these interference effects are their {\it shape}---
the characteristic field or energy to which they are sensitive, or more
precisely the functional dependence on these parameters---
and their {\it magnitude}--- simply how big these quantum corrections are
compared to the classical conductance.
Here we concentrate on the magnitude and merely note that
a theory for the shape has been extensively
developed.\cite{Blu88,JalBarSto,Pluhar94,Efetov94}

A theory of the magnitude of quantum interference effects in chaotic
cavities has recently been developed by making a statistical ansatz for the
$S$-matrix describing the
scattering.\cite{Pluhar94,Efetov94,Iid90,harold-pier,rodolfoEPL94}
Refs. \onlinecite{harold-pier} and \onlinecite{rodolfoEPL94} developed a
random $S$-matrix theory by assigning to $S$ an ``equal a priori distribution''
once the symmetry requirements were imposed. The results for the average,
variance, and probability density of the conductance were in good
agreement with numerical calculations for a
chaotic cavity connected to two waveguides.\cite{harold-pier} However, the
random-matrix predictions for both the weak-localization correction
and the variance are larger than the experimental results.\cite{Chang94,chan}
In addition, the measured probability density is close to a Gaussian
distribution\cite{chan} when there are two propagating modes per lead ($N=2$),
while random-matrix theory predicts a Gaussian distribution only for
$N \geq 3$.\cite{harold-pier}

Inherent in the model of Refs. \onlinecite{harold-pier} and
\onlinecite{rodolfoEPL94} is the assumption (among others) that one can
neglect phase-breaking processes which destroy the coherence of the wave
function. In this paper we show that this assumption is largely responsible
for the discrepancy between theory and experiment mentioned above.
We make specific predictions for the dependence of the quantum transport
corrections on the degree of phase-breaking which may be tested by future
experiments.

To simulate the effects of phase-breaking events we adopt a model
suggested by M. B\"{u}ttiker:\cite{buttiker} in addition to the
physical leads $1,2$ attached to reservoirs at chemical potentials
$\mu _1, \mu_2$, a lead $3$  connects the cavity to a
phase-randomizing reservoir at $\mu  _3$.
This model has been discussed extensively for disordered
materials\cite{phasebreak} and, more recently, for
ballistic quantum dots by Marcus, {\it et al.}.\cite{Mar93}
A similar model has been used for absorption of microwaves in
chaotic-scattering from cavities.\cite{Dor91} Requiring
the current in lead $3$ to vanish determines $\mu _3$; the two-terminal
dimensionless conductance is then found to be
\begin{equation}
\label{buttiker}
g \equiv G/(e^2 /h) = 2\left[ T_{21} +
 \frac {T_{23}T_{31}}{T_{32}+T_{31}}\right] \; \; ,
\end{equation}
where $T_{ij}$ is the transmission coefficient for ``spinless
electrons'' from lead $j$ to lead $i$. The factor of $2$ accounts for spin
explicitly.
We call $N$ the number of channels in leads $1$ and $2$, $N_{\phi}$ that
in lead $3$, and $N_T=2N+N_{\phi }$.

We now make the fundamental assumption of an ``equal a priori distribution''
for the {\it total} $N_T \times N_T$ scattering matrix $S$,
once the symmetry requirements have been imposed. $S$ is, of course,
unitary, and is symmetric in the absence of a magnetic field because
of time-reversal symmetry. We assume that the statistics of
the total $S$-matrix are given by the Circular Ensembles
of random matrix theory.\cite{ReviewRMT} For $B=0$ the orthogonal ensemble
(COE, denoted $\beta = 1$) is appropriate while for nonzero $B$ we use the
unitary ensemble (CUE, $\beta = 2$). In contrast to previous studies of the
eigenphases,\cite{Blu88,rodolfoEPL94}
total transmission,\cite{harold-pier,rodolfoEPL94}
or individual $S$-matrix elements,\cite{Dor91}
we treat the statistics of $g$ given in Eq. (\ref{buttiker}).
In the rest of the paper, we first derive results valid in the weak
and strong phase-breaking limits, then combine these into
an interpolation formula which simulations show to be
valid, and finally compare with experiments.

We start by recalling the result for the WLC and variance at
$N_{\phi}=0$ given in Ref. \onlinecite{harold-pier}:
\begin{mathletters}
\label{varwlN}
\begin{equation}
\delta g \equiv  \langle g \rangle ^{(\beta =1)} -
\langle g \rangle ^{(\beta =2)}
= -N/(2N+1)
\label{wlN}
\end{equation}
\begin{equation}
{\rm var}g =
\left\{
\begin{array}{ll}
\parbox{1.5in}{ \vspace{-\abovedisplayskip} \[
\frac{ 4N(N+1)^2 }{ (2N+1)^2(2N+3) } \; ,
\] \vspace{-\belowdisplayskip} }
& \beta = 1 \\
\parbox{0.8in}{ \vspace{-\abovedisplayskip} \[
\frac{ N^2 }{ 4N^2-1 } \; ,
\] \vspace{-\belowdisplayskip} }
& \beta = 2 \; .
\end{array}
\right.
\label{varN}
\end{equation}
\end{mathletters}
In addition to these results, the probability density
of $g$ was calculated for $N=1-3$.

The case $N_{\phi}=1$ and $\beta =1$ is a special one which can be analyzed
in detail.
{}From the joint probability distribution of $S$-matrix elements in one row,
\cite{pereyra-mello}
\begin{equation}
P(S_{11},\ldots,S_{1N_T}) \propto
(1 - | S_{11} |^2 )^{-N} \delta( 1 - \sum_{i=1}^{N_T} | S_{1i} |^2 )
\; ,
\end{equation}
one can show that the joint distribution of the $T_{3j}$'s is
\begin{equation}
P(\{T_{3j}\}) =
\frac{ N(2N-1)! (T_{31}T_{32})^{N-1} \delta(1 - \sum T_{3j}) }
{ [(N-1)!]^2 (T_{31} + T_{32})^N } \; .
\end{equation}
Remarkably, the $ \langle g \rangle $ that one obtains by integrating over
this distribution is identical to that for $N_{\phi} = 0$.

In the limit $N_{\phi} \gg 1$, one can obtain a number of results
to leading order in $1/N_{\phi}$. First, expand the
conductance of Eq. (\ref{buttiker}) in powers of $\delta T_{ij}$ where
$T_{ij}= \langle T_{ij} \rangle + \delta T_{ij}$.
By definition $\langle \delta T_{ij} \rangle =0$; using the methods of Ref.
\onlinecite{melloJPA90}, one finds that the correlations among the $T_{ij}$'s
are given by
\begin{mathletters}
\label{Tcorrel}
\begin{eqnarray}
\lefteqn{ \langle  \delta T_{ij} \delta T_{kl} \rangle = A_{\beta} \left\{
N_i N_j (N_T+\delta_{\beta 1})(N_T+2\delta_{\beta 1})\delta_{ik}\delta_{jl}
\right. } \hspace{0.2in} \nonumber
\\
 & & \mbox{} - N_i N_j N_l (N_T+\delta_{\beta 1})\delta_{ik}
- N_i N_j N_k (N_T+\delta_{\beta 1})\delta_{jl} \nonumber
\\
 & & \left. \mbox{} + N_i N_j N_k N_l ( 1 + \delta_{\beta 1}) \right\} \; ,
\end{eqnarray}
\begin{equation}
A_1 = [N_T (N_T+1)^2 (N_T+3) ]^{-1} \; ,
\end{equation}
\begin{equation}
A_2 = [N_T^2 (N_T^2 - 1) ]^{-1}
\end{equation}
\end{mathletters}
for $i \neq j$, $k \neq l$.
(For $\beta = 1$, the indices of the $T_{ij}$'s have
been permuted so as to maximize coincidences.)
Note that $\langle \delta T_{ij} \delta T_{kl} \rangle $ is at least of
order $1/N_{\phi }^2$. Thus
\begin{equation}
\label{<g>}  \langle g \rangle = 2\left[  \langle T_{21}
\rangle +\frac{ \langle T_{23} \rangle  \langle T_{31} \rangle }
{ \langle T_{32} \rangle + \langle T_{31} \rangle }\right]  +
O (1/N_{\phi }^2).
\end{equation}
Here we have\cite{harold-pier}
$ \langle T_{ij} \rangle =$ $N_i N_j /(N_T +\delta _{\beta 1})$, a
result closely related to the Hauser-Feshbach formula in nuclear
physics.\cite{bill} The WLC is then given by
\begin{equation}
\label{WLC}\delta g = - N/N_{\phi} + O ( 1/N_{\phi}^2 ) .
\end{equation}
For the variance one finds from Eq. (\ref{Tcorrel})
\begin{equation}
\label{varg}{\rm var}g=\left( \frac{N}{N_{\phi}}\right) ^2
\frac{2}{\beta }\left[ 1+\frac {2-\beta }{2N}\right] + \cdots
\end{equation}
and, for the ratio of the variances for $\beta =1,2$,
\begin{equation}
\label{ratiovarg}\frac{({\rm var}g)^{(\beta =1)}}
{({\rm var}g)^{(\beta =2)}} = 2\left( 1+\frac{1}{2N}\right)
+ \cdots \; .
\end{equation}
Of course the magnitude of the quantum corrections decreases as the number
of phase-breaking channels increases. Note that the power controlling the
decay of the variance is twice that of the average conductance.
The deviation of the ratio of the variances from 2 is highly unusual;
in fact, for $N=1$ the ratio can be as high as 3.

We present in Figs. 1, 2 and 3 results of simulations of the random-matrix
model in order to check the asymptotic behavior and suggest improvements.
The points are obtained by generating random  $N_T \times N_T$ unitary
or orthogonal matrices and computing $g$ from Eq. (\ref{buttiker}).
In Fig. 1, the value $N=2$ was selected for comparison with
the experiment of Ref. \onlinecite{chan}.
The log-log insets in Fig. 1 show that the convergence of the numerical
to the asymptotic results [Eqs. (\ref{WLC})-(\ref{varg}), dotted lines]
is rather slow.
The curves in the main parts of Fig. 1 are interpolation formulae suggested
by the results for $N_{\phi }=0$ and $N_{\phi } \gg 1$. For the WLC,
Eqs. (\ref{wlN}) and (\ref{WLC}) suggest the relation
\begin{equation}
\label{interpWLC}
\delta g \simeq  -N/(2N+N_{\phi }) \; .
\end{equation}
For the fluctuations of the conductance, we combine square roots of variances,
\begin{equation}
\label{interprmsg}
({\rm var}g)^{1/2}=\left[ ({\rm var}g)^{-1/2}
_{N_{\phi }=0} +({\rm var}g)^{-1/2}_{N_{\phi } \gg 1} \right]^{-1} \; ,
\end{equation}
because in the numerical
simulations the change in ${\rm var}g$ is clearly linear in $N_{\phi }$,
not quadratic. So, for $\beta =2$
\begin{equation}
\label{interpvarg}
{\rm var}g \simeq N^2 / [ (4N^2 -1)^{1/2}+N_{\phi } ]^2 \; .
\end{equation}
For $\beta =1$ a similar, but more complicated, expression holds.
These interpolation formulae agree very well with the numerical results;
; the only significant deviation is for $N=1$ and small $N_{\phi}$
[Fig. 1(c)].

The results at fixed values of $N_{\phi}$ shown in Fig. 2 are relevant to
experiments at fixed temperature in which the size of the
opening to the cavity is varied. Though
$\delta g$ and ${\rm var} g$ are nearly independent of $N$ in the
perfectly coherent limit--- a result known as ``universality''---
phase-breaking channels cause the magnitudes
to vary. Thus the universality can only be seen if
$N_{\phi} \ll N$; otherwise, the behavior is approximately linear, as in
some experiments.\cite{MJBer94,chan} Clearly, phase-breaking must be
included in interpreting the experiments.

In addition to the mean and variance, the probability density of the
conductance, $w (g)$, is experimentally measurable.\cite{chan}
The numerical results in Fig. 3 show that as $N_{\phi}$ increases $w(g)$ tends
towards a Gaussian and is therefore fully characterized by the mean and
variance given above. For $N \geq 3$ the distribution is essentially
Gaussian\cite{harold-pier} even for $N_{\phi} = 0$, so that $N_{\phi}$ affects
$w(g)$, apart from changing the mean and variance, only for $N < 3$.
For $N=1$, the highly non-Gaussian distributions\cite{harold-pier,rodolfoEPL94}
at $N_{\phi} = 0$  become approximately Gaussian
at $N_{\phi}=3$; the intermediate distributions, $N_{\phi} = 1,2$, are shown
in Fig. 3. For $N=2$, the deviation of
$w(g)$ for $N_{\phi} =0$ from a Gaussian is smaller than for $N=1$, and
hence fewer phase-breaking channels produce a Gaussian distribution.
We show numerical results only for $N_{\phi} = 1$.

Finally, we use our model to interpret the experiments
of Refs. \onlinecite{MJBer94,Chang94} and \onlinecite{chan}.
First, the solid circles in Fig. 1 show the measured $N_{\phi}$, $\delta g$,
and ${\rm var}g$ of Ref. \onlinecite{chan} in the $N=2$ case. Before
comparison the experimental variance must be corrected for thermal averaging:
convolution over the derivative of the Fermi function produces a reduction
factor of $ \sim 0.22 - 0.38$ for an electron temperature of $50 - 100$~mK.
We have increased the measured ${\rm var}g$ by the inverse of this
reduction factor; the WLC is not affected by thermal averaging since it is
already an average effect. The error bars shown result from both the
uncertainty in temperature in the case of the variance and the experimental
fluctuations at small $B$ for the variance and WLC.
(For the moment, we do not assign further physical significance to the
oscillations seen in the experiment.)
Note that in Fig. 1 we have not fit the theory
to the data and yet the agreement is very good.  Second, in Fig. 2 we show
as solid symbols the data of all three Refs. \onlinecite{MJBer94,Chang94}
and \onlinecite{chan}. The value of $N_{\phi}$ deduced from comparing
to our calculations is in good agreement with the value estimated
independently in the experiments:
$N_{\phi} = 7-9$ for Ref. \onlinecite{MJBer94},
$N_{\phi} = 2$ for Ref. \onlinecite{Chang94}, and
$N_{\phi} = 4-8$ for Ref. \onlinecite{chan}.
Finally, the probability density of the conductance is found to be
Gaussian in Ref. \onlinecite{chan}. For the estimated
$N_{\phi} = 4 - 8$, we obtain a Gaussian distribution and so are consistent.
Observation of the interesting non-Gaussian distribution of $g$ obtained
in the theory for $N=1,2$ and $N_{\phi} = 0$ requires greatly reduced
phase-breaking.

In conclusion, we have presented a random-matrix model that simulates
the effects of phase-breaking events on electronic transport through
ballistic chaotic cavities. The analysis of recent experiments indicates that
one can find a value of $N_{\phi }$, the number of phase-breaking channels,
that allows a consistent description of the data.
Further experiments are needed to test quantitatively the dependence on
$N$ and $N_\phi$ that we predict.

We thank C. M. Marcus for several valuable conversations.

{\it Note added}--- While preparing this paper for publication, we received
a preprint by P. W. Brouwer and C. W. J. Beenakker with some overlapping
material for $N=1$.

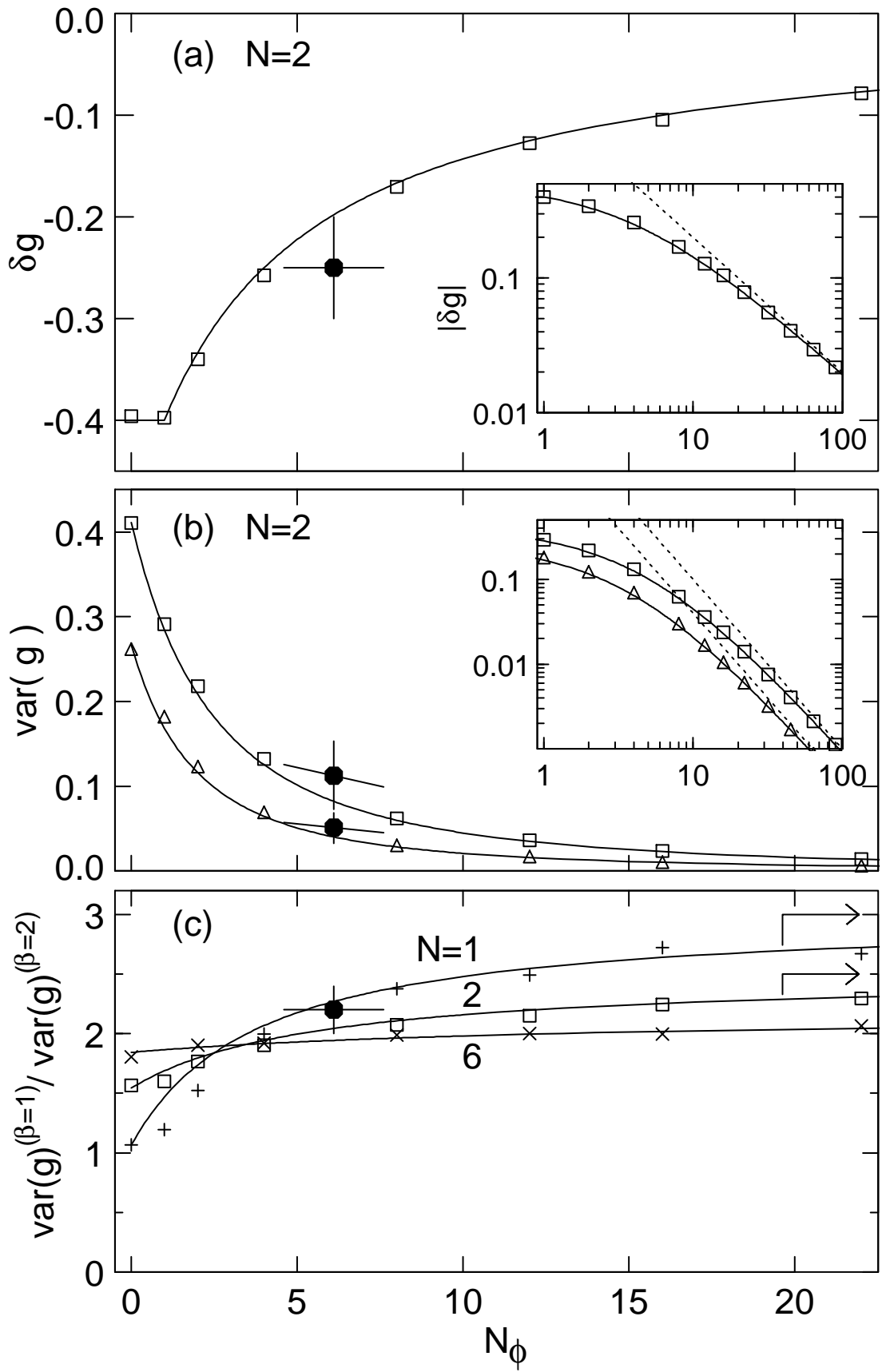
\begin{figure}
\caption{
Magnitude of quantum transport effects as a function of the number of
phase-breaking channels, $N_{\phi}$, on linear (main panels) and log-log
(insets) scales.
(a) The weak-localization correction ($N=2$). (b) The variance in both the
orthogonal (squares) and unitary (triangles) cases ($N=2$). (c) The ratio
of the variance in the orthogonal case to that in the unitary for $N=1$,
$2$, and $6$; the arrows mark the $N_\phi \rightarrow \infty$ limit
for $N=1,2$.
Open symbols  are numerical results ($20,000$
matrices used, statistical error the size of the symbol).
Solid lines are interpolation formulae.
Dotted lines are asymptotic results.
Solid circles are experimental results of Ref. \protect\onlinecite{chan}
corrected for thermal averaging.
The interpolation formulae are excellent except for $N=1$ and small $N_{\phi}$
[panel (c)].
}
\end{figure}

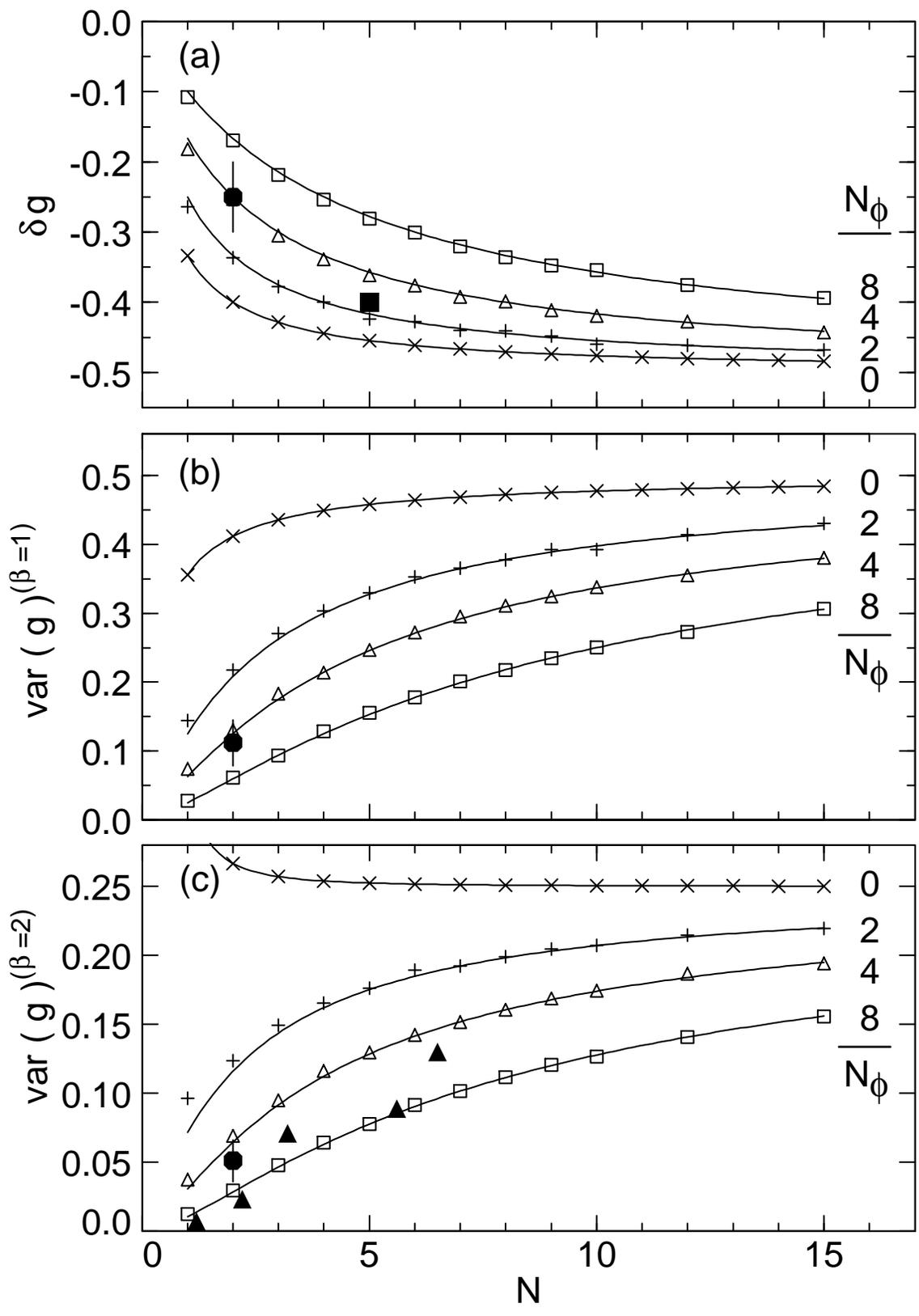
\begin{figure}
\caption{
The magnitude of quantum transport effects as a function of the number of
channels in the leads, $N$, for fixed $N_{\phi} = 0$, $2$, $4$, and $8$.
(a) The weak-localization correction. (b) The variance for the orthogonal
case ($B=0$). (c) The variance for the unitary case (nonzero $B$).
Open symbols  are numerical results (as in Fig. 1).
Solid lines are interpolation formulae.
Solid symbols are experimental results of Refs.
\protect\onlinecite{MJBer94} (triangles),
\protect\onlinecite{Chang94} (squares), and
\protect\onlinecite{chan} (circles) corrected for thermal averaging.
The introduction of phase-breaking decreases the ``universality'' of
the results but leads to good agreement with experiment.
}
\end{figure}

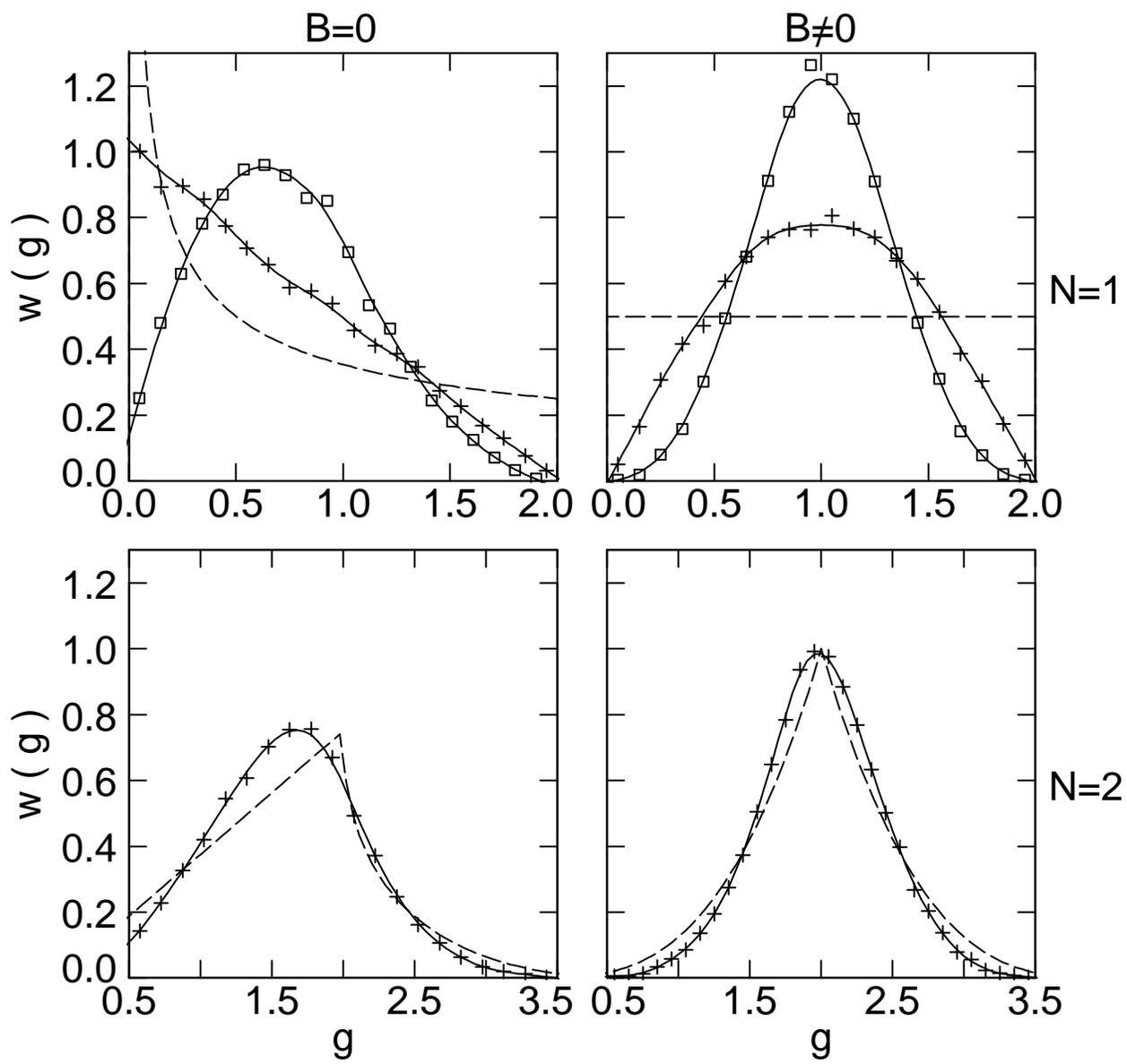
\begin{figure}
\caption{
The probability density of the conductance in the orthogonal (first
column) and unitary (second column) cases for $N=1$ (first row) and $N=2$
(second row).  Increasing the phase-breaking from zero (dashed lines,
analytic) to $N_{\phi} = 1$ (plus symbols, numerical) to $N_{\phi}=2$
(squares, numerical) moves the distribution towards a Gaussian.
}
\end{figure}

\end{document}